\documentstyle[11pt,newpasp,twoside,epsf]{article}
\begin{document}
\title{The  Stability  Analysis  of  the  Exoplanetary  Systems}
\author{Jianghui JI$^{1,2}$,  Lin Liu$^{3}$, H. Kinoshita$^{4}$,
 Guangyu Li$^{1,2}$, H. Nakai$^{4}$}
\affil{1.Purple  Mountain  Observatory, Chinese  Academy  of
Sciences, Nanjing, 210008}

\affil{2.National Astronomical Observatory, Chinese Academy of
Sciences,Beijing 100012,China}

\affil{3. Department of Astronomy, Nanjing University, Nanjing
,210093}

\affil{4.National Astronomical Observatory, Mitaka, Tokyo
181-8588,Japan}

\begin{abstract}
To date,  more than 100 giant Jupiter-like planets have been
discovered in Doppler surveys of solar-type stars.  In this paper,
we perform simulations to investigate three systems: GJ 876, HD
82943 and 55 Cnc. The former two systems both have a pair of
planets in the 2:1 Mean Motion Resonance (MMR), while the inner
two companions of the later is close to 3:1 MMR.  By integrating
hundreds of the planetary orbits of  three systems for million
years, we find that for GJ 876 and HD 82943, the critical argument
$\lambda _{1} - 2\lambda _{2} + \varpi_{1}$  and $\lambda _{1} -
2\lambda _{2} + \varpi_{2}$ librate about $0^{\circ}$ or
$180^{\circ}$, indicating 2:1 MMR can play an important role in
stabilizing the motion of the planets so that they are protected
from frequent close encounters. As  for 55 Cnc,  we further show
the three resonant arguments for 3:1 MMR execute librations for
millions of years respectively, which reveals the evidence of  the
resonance for this system. Additionally, we should  emphasize
another vital mechanism is the apsidal phase-locking between a
couple of planets for a certain system. For GJ 876 and HD 82943,
we discover the relative apsidal longitudes $\varpi_{1} -
\varpi_{2}$ move about $0^{\circ}$ or $180^{\circ}$, respectively;
but for 55 Cnc, we find that there exists an asymmetric apsidal
libration between two inner planets. Finally, we made a brief
discussion about the Habitable Zones in the exoplanetary systems.
\end{abstract}

\section{Introduction}
The discovery of  the extrasolar planets is opening a new world
beyond our solar system.  Ever since 1995,  Mayor \& Queloz (1995)
detected the first extrasolar giant Jupiter--51 Peg and as of
August 5, 2003, there are 111 exoplanets discovered (Butler et al.
2003 and references therein;also see http://exoplanets.org) using
the radial velocity technique in the surveys of the solar-type
stars. Nowadays, a dozen of multiple planet systems--HD 82943, GJ
876, HD 168443, HD 74156, 47 Uma, HD 37124, HD 38529, HD 12661, 55
Cnc ,Ups And  (see Table 8 in Fischer et al. 2003), HD 169830
(Mayor et al. 2003), and HD 160691 (Jones et al. 2003;
Gozdziewski, Konacki \& Maciejewski 2003; Kiseleva-Eggleton et al.
2002) were detected in recent years. And it is necessary to
classify the discovered multiple planetary systems according to
their statistical characteristics (such as planet masses,
semi-major axes, eccentricities and metallicity) (Marcy et al.
2003; Santos et al. 2003;also see Figure 1a-b) and to study the
correlation between mass ratio and period ratio (Mazeh \& Zucker
2003). And the other important point is to investigate the
possible stable configurations for the multiple systems,
especially for the couples in the mean motion resonance.

\begin{figure}
\plottwo{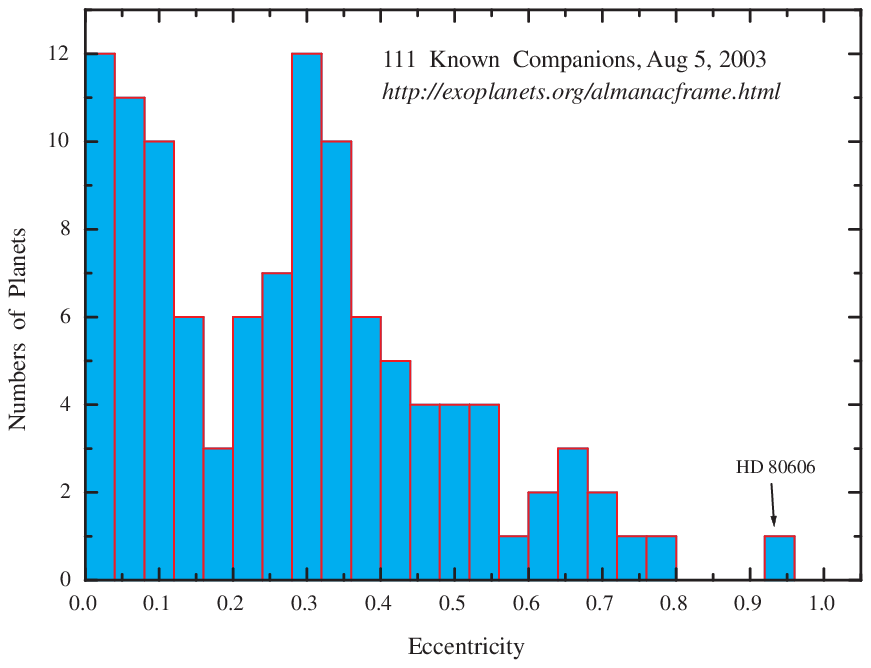}{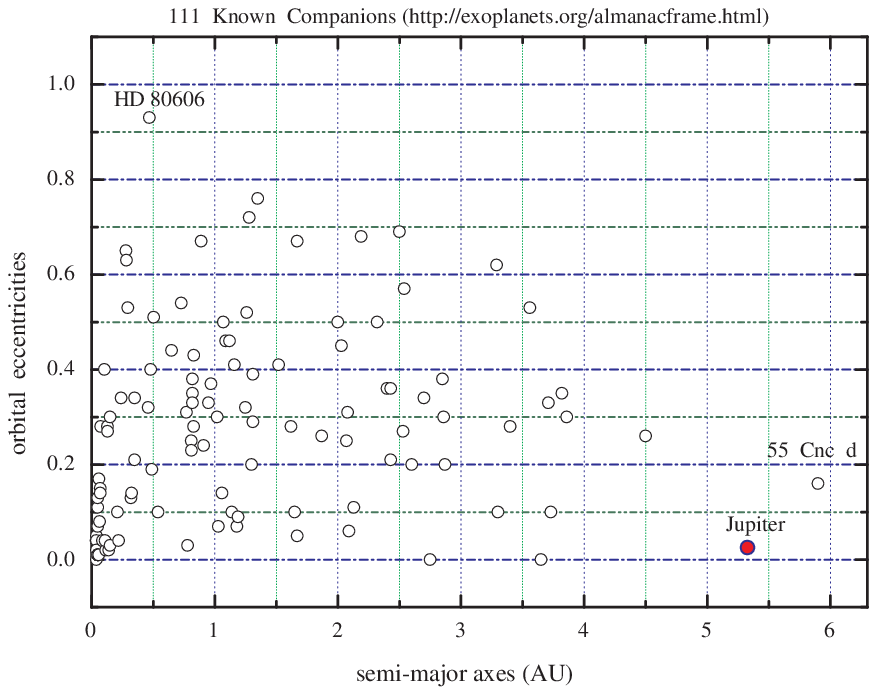} \caption{Left panel (a): The histogram
for the eccentricities for 111 extrasolar planets discovered. The
figure  exhibits  that  more than 50\% of the planets have the
eccentricities  larger  than   0.30,   and HD 80606 b can occupy
the  eccentricity  up  to  0.93,  indicating that they are quite
different  with the counterparts in our solar system. Right panel
(b): The distribution of the semi-major  axes versus the
eccentricities  for  all planets. Note  that  55 Cnc d can be
distant as far as nearly 6 AU from its parent star, which can  be
compared  with Jupiter at a distance of  5.2 AU  from Sun.}
\end{figure}

\begin{table}
\caption  {The  astrocentric orbital parameters for  three systems
in MMR}
\begin{tabular}{lclllc}
\tableline \tableline
Planet  &Mass (Mjup)  &Period (day)  &a (AU)  &ecc.   &MMR \\
\tableline
GJ 876 b    &1.06   &29.995  &0.1294   &0.314   &2:1  \\
GJ 876 c    &3.39  &62.092  &0.2108  &0.051 &2:1\\
HD82943 b   &1.63  &444.6   &1.16    &0.41    &2:1 \\
HD82943 c   &0.88  &221.6   &0.73    &0.54    &2:1 \\
55 Cnc b    &0.83  &14.65 &0.115 &0.03 &3:1  \\
55 Cnc c    &0.20  &44.27 &0.241 &0.41 &3:1\\
55 Cnc d    &3.69  &4780.0 &5.461   &0.28\\
\tableline \tableline
\end{tabular}
\end{table}

\section{ Multiple Planetary Systems in Mean Motion Resonance}
In the multiple planetary systems are increasingly found  the
existence of  the  MMR, such as  HD 82943  (Gozdziewski  \&
Maciejewski 2001),  GJ 876 (Lee  \&  Peale 2002), and possibly HD
160691 (Gozdziewski et al. 2003) in a 2:1 MMR, 55 Cnc in a 3:1 MMR
(Marcy et al. 2002; Ji et al. 2003a). Fischer D.A. (private
communication) pointed out that more than half of the solar-like
stars with one detected planet show velocity trends indicating the
presence of a second or multiple planets. Thus, we turn to the
interesting topic of  studying  the stable geometry for two
planets locking into a  resonance. And our investigations are
helpful in an understanding of the dynamics and possible
mechanisms of sustaining the stability of  such systems. In Table
1\footnote{The data are, respectively, taken from  Laughlin \&
Chambers (2002), http://obswww.unige.ch, Fischer et al. (2003).}
are listed the orbital parameters  of  HD 82943, GJ 876 and 55
Cnc. Another important feature for the resonant systems is the
apsidal lock between the orbiting companions, indicating that the
relative apsidal longitudes of two orbits librate about a
constant, such that two planets have common time-averaged rate of
apsidal precession (Kinoshita \& Nakai 2000; Chiang  \& Murray
2002; Malhotra 2002).

\subsection{HD 82943 and GJ 876}
The critical argument $\theta_{1}$, $\theta_{2}$ for the 2:1 MMR
is
\begin{equation}
\label{eq1} \quad\quad\theta_{1} = \lambda _{1} - 2\lambda _{2} +
\varpi_{1},
\end{equation}
\begin{equation}
\label{eq2} \quad\quad\theta_{2} = \lambda _{1} - 2\lambda _{2} +
\varpi_{2},
\end{equation}

And the relative apsidal longitude is
\begin{equation}
\quad\quad\theta_{3}=\Delta\varpi = \varpi_{1} -  \varpi_{2}.
\end{equation}
\noindent where $\lambda _{1,2}$ are , respectively, the mean
longitudes of the inner and outer planets, and $\varpi_{1,2} $,
respectively, denote their periastron longitudes.

There are three possible stable configurations for a system in a
2:1 MMR :(I) Only $\theta_{1}$ librates about $0^{\circ}$;
(II)Case of alignment,
$\theta_{1}\approx\theta_{2}\approx\theta_{3}\approx 0^{\circ}$
;(III)Case of antialignment, $\theta_{1}\approx 180^{\circ}$,
$\theta_{2}\approx0^{\circ}$, $\theta_{3}\approx180^{\circ}$. The
simulation results  show that HD 82943  could  occupy the above
three  stable configurations, while GJ 876 does Classes I and  II.
Figure 2a displays an antialignment case for HD 82943. The
librations of the relative apsidal longitudes and mean motion
resonant variables about 0 or 180 degrees show that two important
dynamical factors to characterize the multiple planetary systems.

As  for HD 82943, we established a  semi-analytical model on the
averaged Hamiltonian (Ji  et al. 2003b; also see Figure 2b) to
explore the  apsidal  motion  for  this system and found that
although  the  two  planets   have   high eccentricities  to 0.54
and  0.41,  the  system remain  stable   for  10  Myr because of
both dynamical  mechanisms  of  the   2:1   MMR and apsidal lock.
This  is also true  for the HD 160691 system (see Bois et al.
2003, where the stable configuration is involved in a 2:1 mean
motion   resonance with  an anti-aligned in  two apsidal lines,
with higher eccentricity  of the  outer planet).

\begin{figure}
\plottwo{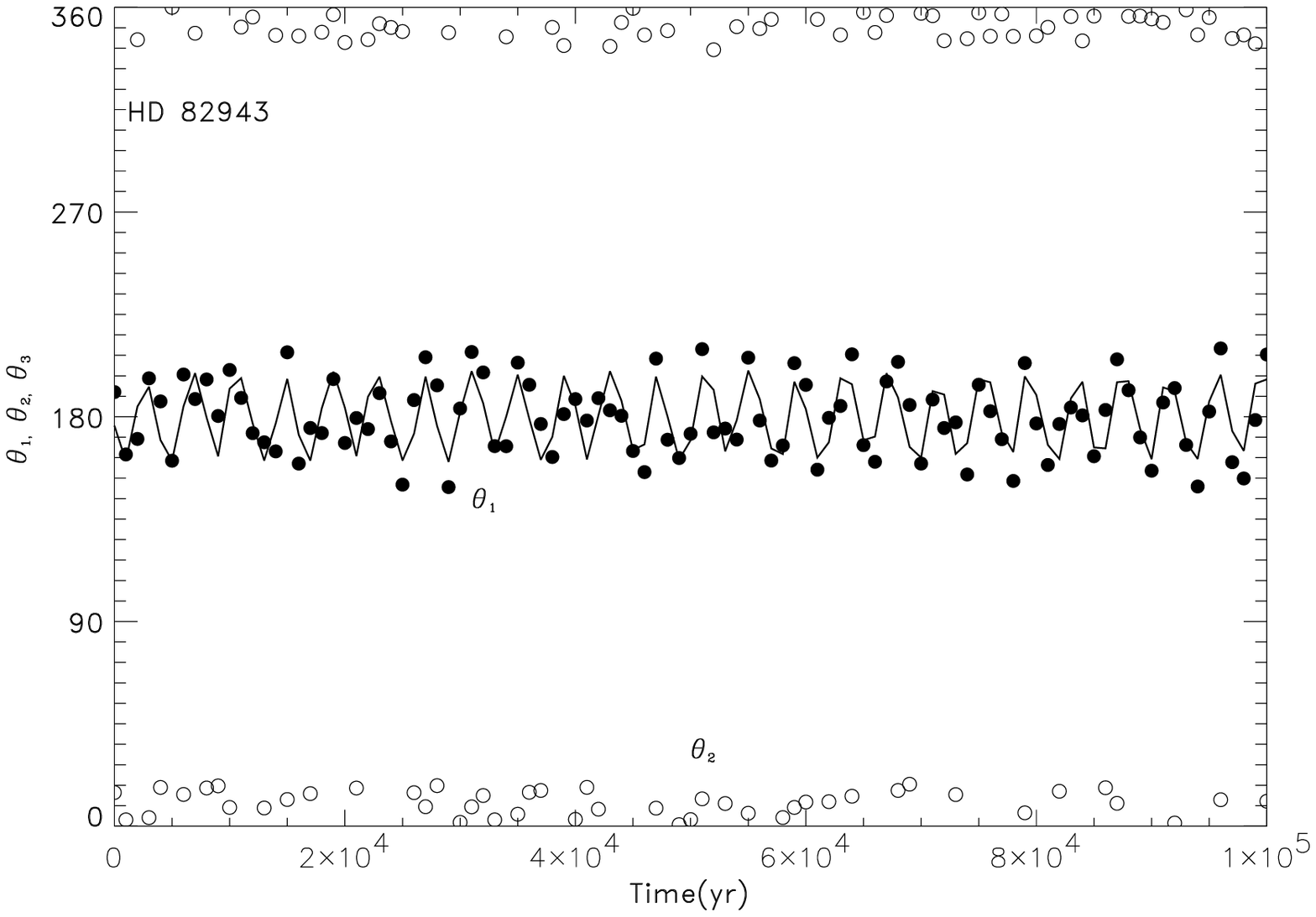}{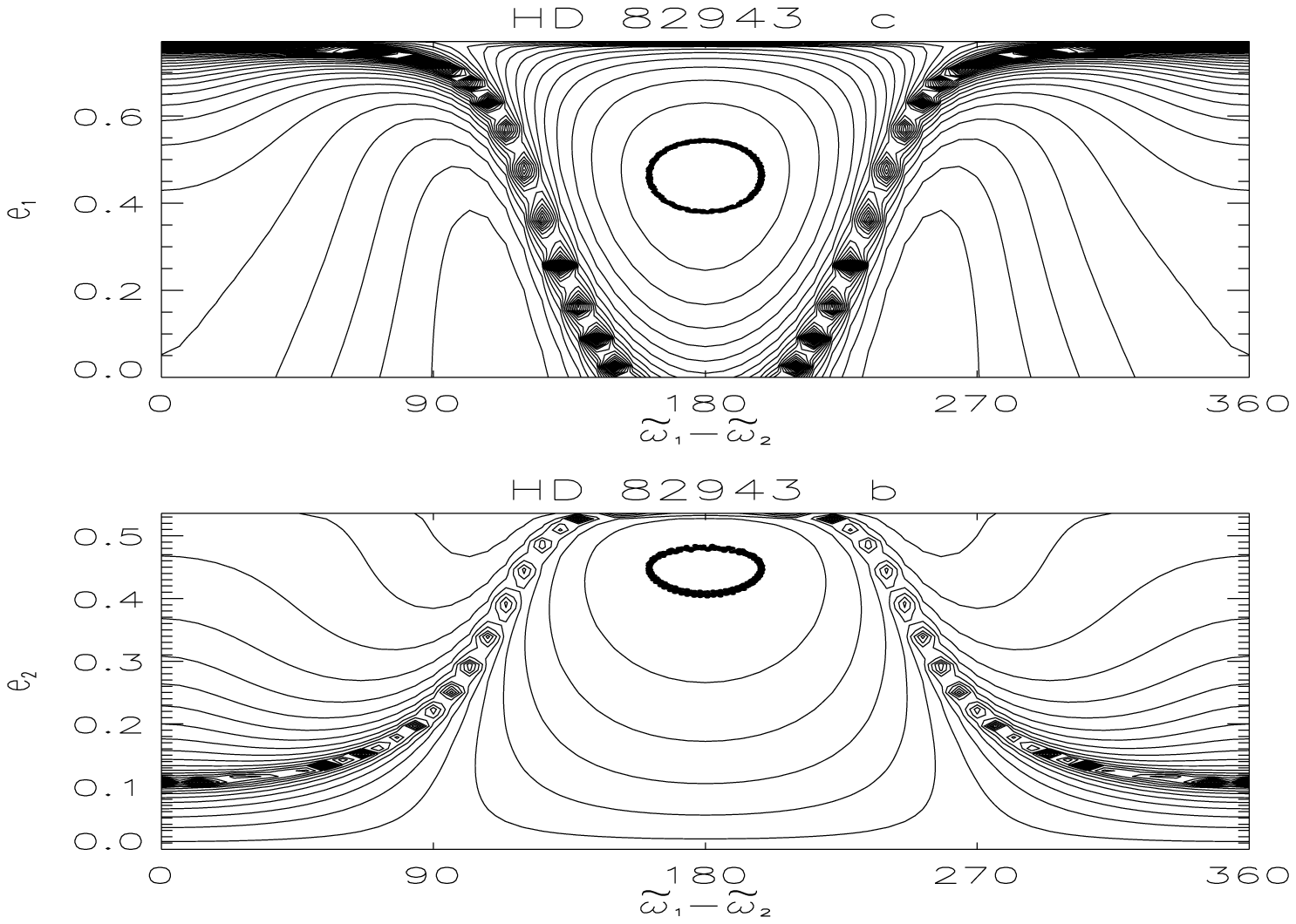} \caption{Left Panel (a): Antialignment
case for HD 82943.  The figure exhibits  that $\theta_1$  (filled
circle) and $\theta_2$ (empty circle)  librate about $180^{\circ}$
and  $0^{\circ}$, respectively, for millions years. The  relative
apsidal longitude of $\theta_3$ (solid line)  move about
$180^{\circ}$. Right panel (b): The eccentricity versus
$\Delta\varpi$ evolution. The theoretical Hamitonian contour (thin
lines) concords with the numerical simulations (thick line).}
\end{figure}

\subsection{55 Cnc}
The work of  Ji  et al.(2003a) confirmed that the inner two
companions of  55 Cnc  (Marcy et al. 2002) are in a  3:1 resonance
and still experience the apsidal phase-locking. Two characterized
arguments for resonances are:
\begin{equation}
\label{eq1} \quad\quad\theta_{1} = \lambda _{1} - 3\lambda _{2} +
2\varpi_{2}, \Delta\varpi = \varpi_{1} -  \varpi_{2}.
\end{equation}
\noindent where $\lambda _{1,2}$ are, respectively, the mean
longitudes of  55 Cnc  b  and  55 Cnc c, and $\varpi_{1,2} $,
individually, denote their apsidal longitudes. In Figure 3a,
$\theta_{1}$ librate about $90^{\circ}$ for million years. The
relative apsidal longitude of $\Delta\varpi$ walks about
$250^{\circ}$, which indicates that there exists an asymmetric
apsidal libration between two inner planets.  Here two dynamical
mechanisms are responsible for the stability of the 55 Cnc system.
In addition, the numerical results further show that a wider
stable region between 55 Cnc c and 55 Cnc d, implying an
Earth-like planet may survive (see \S3) in the triple-planet
system.

\section{Possible Habitable Zones (HZ) in  the  exoplanetary  systems}
Do the terrestrial planets exist in the Habitable Zones (Kasting,
Whitmore, \& Reynolds 1993) for the multiple systems? Is there
possibly a stable liquid water environment for other
civilizations? With the means of the direct integrations, we
performed many tests in 55 Cnc by placing an Earth-mass planet at
$\sim 1$ AU with low eccentricity and found that such orbits can
survive for 10 Myr or even longer (Figure 3b) and there are wider
stable zones for this system. And the stable orbits can be still
reserved even when we gradually increased the mass of the assumed
planet. Similar cases occur in the system of GJ 876 (Ji et al., in
preparation) and other exosystems, especially for the host star
that harbors two well-separated planets.
Although the present ground-based observations set limitations on
the discovery of the Earth-like planets, Space Interferometer
Mission (SIM) may detect our Earth twins in the future task.

Due to the limited numbers of the exosystems involved in low mean
motion resonance at this stage, thus it is too early to conclude
that the planets tend  to be captured into the resonance in the
early planetary formation.  However, the  resonant geometry may
become a suitable dynamical environment related to the stability
of these systems, and future more observations on the exosystems
will elucidate more clues on this.

\begin{figure}
\plottwo{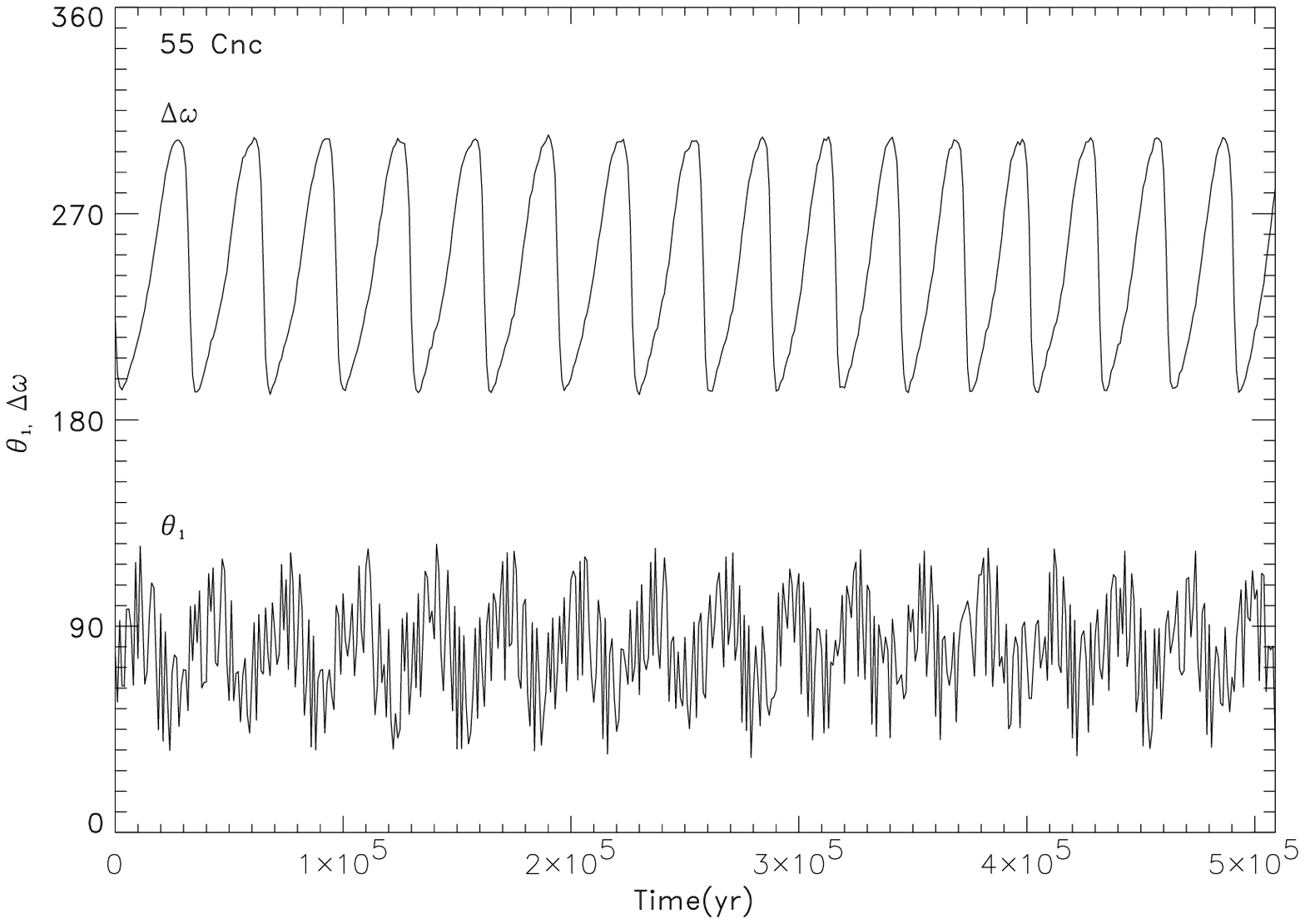}{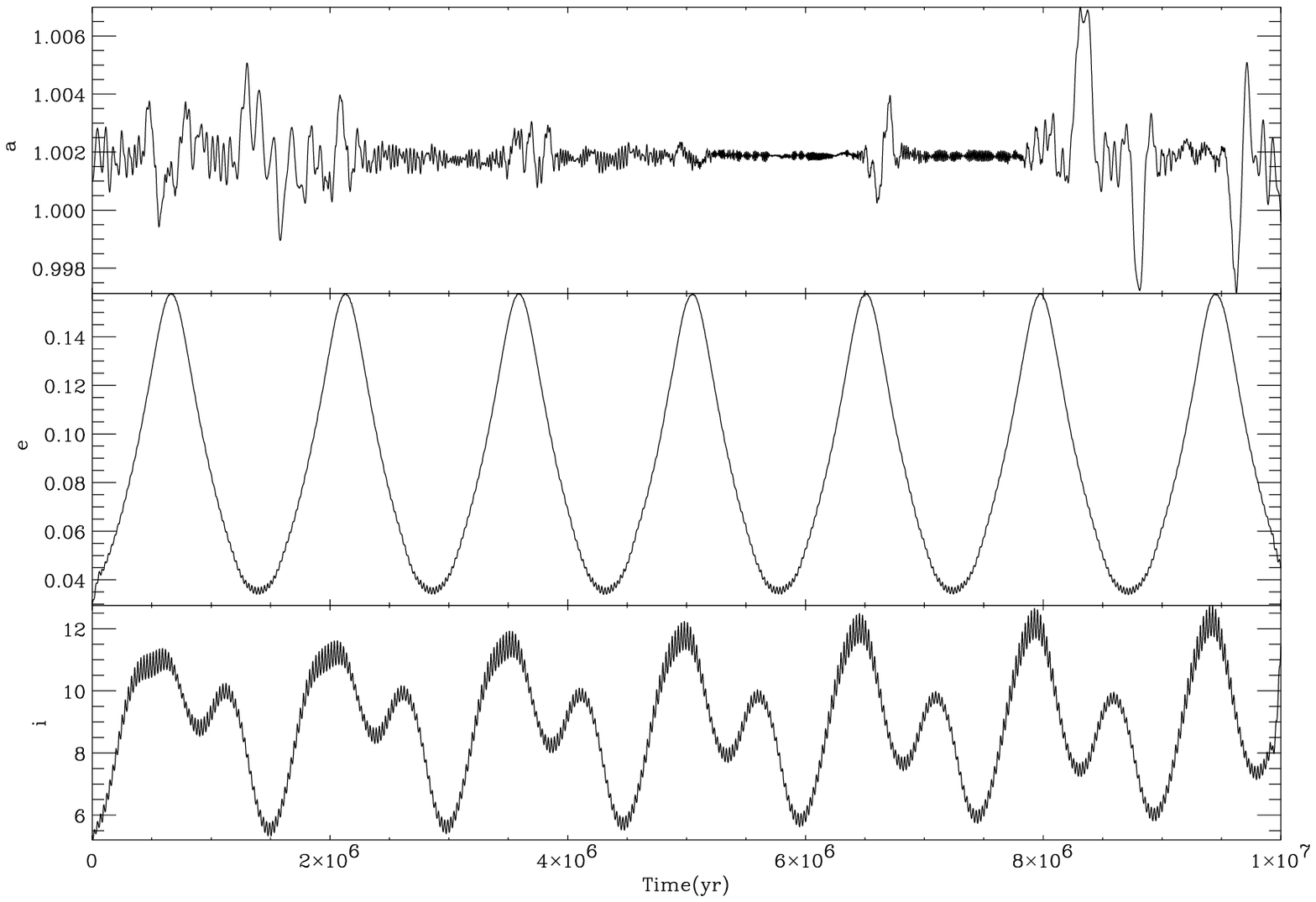} \caption{Left Panel (a): Evidence for a
3:1 resonance for 55 Cnc. The argument of $\theta_{1}$ librate
about  $90^{\circ}$ for million years.The relative apsidal
longitude of $\Delta\varpi$ walks about $250^{\circ}$. Right panel
(b): Orbital evolution of an Earth-mass planet orbiting 55 Cnc at
1 AU. Such orbits may exist and be stable in the planetary
systems.}
\end{figure}

\end{document}